# AN EFFECT OF STIMULATED EADIATION PROCESSES ON RADIO EMISSION FROM EXTENDED SOURCES

Fedor V.Prigara

Institute of Microelectronics and Informatics, Russian Academy of Sciences,

21 Universitetskaya, 150007 Yaroslavl, Russia; fprigara@rambler.ru

ABSTRACT

Both the standard theory of thermal radio emission and the synchrotron theory encounter some difficulties. The most crucial for the former one is nonpossibility to explain the radio spectrum of Venus in the decimeter range (Ksanfomality 1985). The radio spectra of planetary nebulae at high frequencies also are not comfortably consistent with the standard theory (Siodmiak & Tylenda 2001). Here we show that the account for an induced character of radiation processes sufficiently improves the predictions of the standard theory.

Moreover, the developed here theory of radio emission from non-uniform gas gives the radio spectra of extended sources, such as supernova remnants and radio galaxies, which are normally attributed to the synchrotron emission. It is important, in this aspect, that the synchrotron self-absorption produces a change in the polarization position angle across the spectral peak. No such a change was detected in gigahertz-peaked spectrum sources (Mutoh et al. 2002). Besides, the flat or slightly inverted radio spectra of low-luminosity active galactic nuclei are not easily consistent with the synchrotron radiation mechanism (Nagar, Wilson & Falcke 2001, Ulvestad & Ho 2001).

Thermal radiation in a magnetic field is polarized (see, e.g., Lang 1978). This is also true for radiation considered in the present paper, which can be described as incoherent induced radiation (Prigara 2001b).

*Subject headings:* radiation mechanisms: thermal - radio continuum: general

# 1. INTRODUCTION

In this paper the thermal radio emission from non-uniform gas is considered. For example, the gas may be placed in a field of gravity. We suggest, that the thermal radio emission with a wavelength $\lambda$ is emitted by the gaseous layer, for which the next condition of radio emission is valid:

$$\lambda = l = \frac{1}{n\sigma}. \qquad (1)$$

Here $l$ is the mean free path of photons, $n$ is the number density of emitting particles (atoms, ions or molecules), $\sigma$ is the photo absorption cross-section.

It is supposed also, that the thermal radiation is emitted only in the direction, in which the concentration of emitting particles decreases, but not in the opposite direction. The intensity of thermal radio emission is given by Rayleigh-Jeans formula

$$I_\nu = 2kT \frac{\nu^2}{c^2}, \qquad (2)$$

where $T$ is the temperature of emitting gas, $k$ is the Boltzmann constant, $c$ is the velocity of light and $\nu = c/\lambda$ is the frequency of radio emission (e.g., Lang 1978).

It is not supposed, however, that an emitting gaseous layer absorbs all the incident radiation with the same frequency as that of emitted radiation. On the contrary, an emitting gaseous layer may be quite transparent for the emission with frequency $\nu$. We notice, in this context, that the Kirchhoff's consideration (Jammer 1967), which establishes the relation between the emission and absorption coefficients, is not valid for non-uniform systems. In the case of non-uniform system, let us say a planet with its atmosphere, the radiation with given frequency $\nu$ may be absorbed by one part of the system (for instance, by the surface of planet) and emitted by the other part (for instance, by the definite gaseous layer in the atmosphere of planet). The energy transfer from one part of system to the other is realized by the means of convection, thermal conductivity or emission in the other range of frequencies. In this case Kirchhoff's law is valid for the non-uniform system in whole (a planet with its atmosphere), but not for its parts.

## 2. INDUCED EMISSION

In order to elucidate the condition for emission (1) let us consider the Einstein's coefficients $A_{21}$, $B_{21}$ and $B_{12}$, where $A_{21}$ and $B_{21}\rho$ are the numbers of transitions from the energy level 2 to the energy level 1 per unit time, corresponding to a spontaneous and an induced emission respectively, $\rho$ is the energy density of blackbody radiation, and $B_{12}\rho$ is the number of transitions from the energy level 1 to the energy level 2 per unit time, the energy of level 2 being higher than the energy of level 1 ( Jammer 1967; Singer 1959 ) .

In the case of non-degenerated levels the Einstein's coefficients obey the relation

$$\exp(\hbar\omega/kT) = (A_{21} + B_{21}\rho)/(B_{12}\rho), \qquad (3)$$

where $\hbar$ is Planck's constant, $\omega$ is the circular frequency of light, $T$ is the temperature, and $k$ is Boltzmann's constant. Taking into account the relation $B_{12}=B_{21}$, we obtain from the equation (3)

$$A_{21}/(B_{21}\rho) = \exp(\hbar\omega/kT) - 1. \qquad (4)$$

If $\hbar\omega << kT$ which is the case for thermal radio emission, then $A_{21} << B_{21}\rho$. Therefore the contribution of a spontaneous emission to thermal radiation in radio wavelength range may be neglected . Thus thermal radio emission is produced by induced emission, e.i. the emission of maser type .

Returning to the equation (1) we can interpret the mean free path $l$ as the size of a molecular resonator. The emitting gaseous layer contains many molecular emitters of size $l$. Each molecular emitter produces the coherent radiation, thermal radiation being incoherent sum of radiation produced by individual emitters. (For more details see Prigara 2001a).

## 3. RADIO EMISSION FROM PLANETARY ATMOSPHERES

Let us apply the above-formulated condition for emission to the atmosphere of planet. In this case the concentration of molecules $n$ continuously decreases with the increase of height $h$ in accordance with barometrical formula

$$n = n_0 \exp(-\frac{mgh}{kT}), \qquad (5)$$

where $m$ is the mass of molecule, $g$ is the gravity acceleration; $k$ is the Boltzmann constant (e.g., Chamberlain 1978). Here the temperature of atmosphere $T$ is supposed to be constant. In fact the temperature changes with the increase of height, so the formula (5) describes the change of molecule concentration in the limits of layer, for which we can consider the temperature to be approximately constant (the temperature changes with the increase of height essentially slower than the concentration of molecules).

If the emitting particles are the molecules of definite sort, then their concentration is monotonously decreasing with the increase of height. Therefore, according to the condition for emission (1), the radio waves with the wavelength λ are emitted by the gaseous layer, located at well-defined height *h* in atmosphere. Thus, the relation between the brightness temperature of radio emission and wavelength $T(\lambda)$ reproduces (partially or in whole) the temperature section across the atmosphere *T(h)*.

This is the case for the Venus atmosphere. Here the emitting particles are the molecules of $CO_2$. This oxide of carbon is the main component of Venus atmosphere, its contents being 97%. The data concerning the brightness temperature of radio emission from Venus are summarized in Mayer (1970).

Assuming $\sigma \approx 10^{-15}$ cm$^2$ and using the condition for emission (1), we can find the concentration of emitting molecules *n* in the gaseous layer, which emits the radio waves with given wavelength λ. Then, with the help of formula (5) and the data upon the pressure and the temperature in the lower layer of Venus atmosphere (Chamberlain 1978), we can establish the height of emitting layer in the atmosphere. This procedure gives the temperature section of Venus atmosphere shown in Figure 1. It is not in contradiction with the data received by the means of spacecrafts Venera, since on these apparatus the temperature was measured only in the limits of troposphere, up to the height 55 km.

We observe, that the temperature structure of Venus atmosphere is similar to those of Earth's atmosphere (Chamberlain 1978). Because of this analogy to denote the various layers of Venus atmosphere the same terms were used here as those which are usually applied for the Earth's atmosphere. Up to date there was no acceptable explanation for the decrease of brightness temperature in the decimeter range, corresponding in our consideration to the emission of mesopause region.

Though the direct measurements of a temperature profile at the heights of 100 to 160 km are absent, the electron density profile for these heights is available (Ksanfomality 1985). Using the theory of thermal ionization (Landau & Lifshitz 1976), one can see that the night profile of electron density is in agreement with the temperature profile derived in the present paper.

Quite similarly the brightness temperature of Jovian thermal radio emission in the range of 0.1 cm to 4 cm as a function of wavelength reproduces the temperature section of Jovian atmosphere. See Hubbard (1984), where the data upon the microwave emission of Jupiter are represented, and also Lang (1978); Mayer (1970). In this case the temperature structure of atmosphere also is similar to those of Earth's atmosphere: there are two minimums of temperature (tropopause and mesopause) and one intermediate maximum (mesopeak).

The observational data concerning the thermal radio emission of Mars, Saturn, Uranus and Neptune (Lang 1978) are not so complete as in the case of Venus and Jupiter. These data, however, are in agreement with Earth type temperature structure of atmosphere. In the case of Mars the measurements of pressure and temperature in the lower layer of atmosphere are available (Chamberlain 1978), so one can reproduce using the radio emission data the temperature section of Mars atmosphere. Here the emitting particles are the $CO_2$ molecules as in the case of Venus atmosphere.

The intermediate maximum of temperature in the region of mesopeak perhaps can be explained by absorption and next reemission of infrared radiation transferred from the lower layers of atmosphere (Chamberlain 1978).

## 4. THE GASEOUS DISK MODEL

Consider now a gaseous disk with thickness $d$ and radius $R$, in which the regular radial convection is realized: in the medial plane of disk gas flows to the center of disk and near the upper and lower surfaces of disk returns to periphery. The velocities of convection flows lie in planes, containing the normal to disk plane.

Since the total number of particles is conserved, the bulk concentration of particles (ions) $n$ decreases with the increase of radius $r$ as

$$n \propto r^{-1} . \qquad (6)$$

At the upper and lower surfaces of disk and also at $r=R$ the concentration of particles (ions) decreases to some small value.

Consider now the thermal radio emission of such a disk in the direction of normal to the disk plane. According to the emission condition (1), radio waves with frequency ν are emitted by the region of disk with $r \leq r_\nu$, where $r_\nu$ is the value of radius, for which the bulk concentration of ions $n$ satisfies the relation (1). The spectral density of radio emission flux at frequency $\nu$ is given by formula

$$F_\nu = \pi I_\nu \varphi_\nu^2 = \pi I_\nu \frac{r_\nu^2}{a^2}, \qquad (7)$$

where the intensity $I_\nu$ is given by Rayleigh-Jeans formula (2) and $a$ is the distance from gaseous disk (nebula) to the detector of radio emission.

From relations (1) and (6) we obtain

$$r_\nu \propto \lambda \propto \nu^{-1}. \qquad (8)$$

Assuming the temperature $T$ to be constant in the whole volume of gaseous disk, from relation (7) we find that if $r_\nu < R$, and then the flux $F_\nu$ is independent of frequency: $F_\nu$=const. And if $r_\nu = R$ (that corresponds to the long wave range $\nu<\nu_0$), then the flux $F_\nu$ is proportional to the square of frequency:

$$F_\nu \propto I_\nu \propto \nu^2. \qquad (9)$$

Exactly such spectra are observed for the most of planetary nebulae (Pottasch 1984). For example, in the case of planetary nebula NGC 6543 the turnover wavelength dividing the regions with spectra $F_\nu \propto \nu^2$ and $F_\nu = const$, is $\lambda_0$=30 cm, for NGC 7027 $\lambda_0$=10 cm and for IC 418 $\lambda_0$=20 cm.

The spectrum $F_\nu \propto \nu$ observed for the series of compact nebulae, e.g. Vy 2-2, can be explained by the decrease of temperature with the increase of radius in accordance to the law

$$T \propto r^{-1}. \qquad (10)$$

Planetary nebulae are as a rule stationary, not expanding objects. A motion of details, which was interpreted as the consequence of expansion of nebula, one can attribute to the radial convection. The interpretation of spectral lines splitting as the consequence of expansion leads to some difficulties (Pottasch 1984). It is worthwhile to remark that the existing theory of Doppler effect is likely to be unable to explain the non-uniform shift of spectral lines which is often observed, e.g. in the spectra of early type stars (Ebbets 1980).

The recent observations show the jet structure of some planetary nebulae (Lucas, Cox, & Huggins 2000). The jets can be interpreted as a result of interaction of convective flows with two-dimensional magnetic field of planetary nebula. The last forms and stabilizes the gaseous disk discussed above. The radio image of planetary nebula Hb 12 obtained at the wavelength λ=6 cm (Pottasch 1984,Ch.1) shows the extended details following the force lines of magnetic field for two -dimensional dipole.

## 5. THE WAVELENGTH DEPENDENCE OF RADIO SOURCE SIZE

It follows from equation (8) that the angular size of radio source depends on the wavelength of radio emission, namely the angular dimensions of radio source are proportional to the wavelength λ. The wavelength dependence of radio source size is confirmed by observations (Sharov 1982; Lo et al. 1993; Lo 1982). For instance, in the case of galaxy M31 the radius of the central radio core is 3.5 arcmin at the frequency 408 MHz (λ=74cm) and less than 1.3 arcmin at the frequency 1407 MHz (λ=21 cm) (Sharov 1982). The $\lambda^2$ dependence of source size observed in the case of Sgr A* (Lo et al. 1993; Lo 1982) can be obtained if the effect of gravitational field on the flow of ionized gas is taken into account. At small values of radius this effect changes the equation (6) for equation

$$n \propto r^{-1/2} \tag{11}$$

(Prigara 2002). The radio source Sgr A* has a flat spectrum, corresponding to the relation (10), since in this case $r_\nu \propto \nu^{-2}$.

## 6. RADIO EMISSION FROM SUPERNOVA REMNANTS

In the case of supernova remnants we also can apply the theory of thermal radio emission for gaseous disk. The main difference is that instead of the condition of temperature constancy here often the condition of gas pressure constancy is valid

$$P=nkT=const, \tag{12}$$

where *n* as above is the number of particles (ions) per unite volume, *P* is the pressure and *T* is the temperature of gas, and *k* is the Boltzmann's constant. The condition *P=const* and relation (4) lead to the dependence of gas temperature upon radius *r*:

$$T \propto r. \qquad (13)$$

In the consequence of this relation the flux density of radio emission depends on frequency ν as follows

$$F_\nu \propto T\nu^2 r_\nu^2 \propto \nu^{-1}. \qquad (14)$$

The observed spectra of radio emission for supernovae remnants can be interpreted as the combination of spectra $F_\nu$=const and $F_\nu \propto \nu^{-1}$.

Since the intensity of thermal radio emission is proportional to temperature, the increase of temperature in accordance to relation (13) will cause the enlargement of radio brightness on the edge of nebula. Such effect is observed for the most of supernovae remnants, in particular for Cassiopeia A (Kulkarni & Frail 1993; Shklovsky 1984). However there are supernova remnants with the more uniform distribution of radio brightness and respectfully with more flattened spectra $F_\nu \approx const$ corresponding the condition $T \approx const$ (Kulkarni & Frail 1993; Shklovsky 1984). Notice, that the correlation between the spectral index and the radial distribution of radio brightness is in contradiction with the synchrotron theory of radio emission for supernova remnants.

### 7. THE CLASSIFICATION OF RADIO SOURCES

The theory of thermal radio emission for the gaseous disk describes also radio emission from radio galaxies and quasars. As well as for supernova remnants here the spectra $F_\nu \propto \nu^{-1}$ and $F = const$ are observed. For example, the radio spectrum of Virgo A is the combination of these two spectra, similarly to the spectrum of Cassiopeia A. For the quasar NRAO 530 $T \approx const$, $F_\nu \approx const$, and $r_\lambda \propto \lambda$ in the range of wavelengths of 0.3 cm to 6.3 cm (Bower & Backer 1998).

The spectrum $F_\nu \propto \nu^{-1}$ is typical for extended radio sources with jets (Nagar et al. 2002). Compact radio sources with angular dimensions smaller than 0.1 arcmin as a rule demonstrate the spectrum $F_\nu \approx const$ (Tadhunter et al 2001). The last spectrum is also typical for BL Lac objects. It is worthwhile to notice that theory leading to the law (14) is still valid in the case of the sector of disk (instead of whole disk), so one can apply this theory to the jets.

Thus there are two main types of extended radio sources. I type radio sources are characterized by the stationary convection in gravitational or magnetic field with approximately uniform distribution of gas temperature. II type radio sources have the outflows of gas with approximately uniform distribution of gas pressure.

I type radio sources include planetary nebulae, supernova remnants with pulsars (center-brightened shells) (Kulkarni & Frail 1993) and compact sources corresponding to galactic nuclei. II type radio sources include the most of supernova remnants (edge-brightened shells) (Kulkarni & Frail 1993) and active galactic nuclei with jets.

It is possible that the radio source core belongs to the I type radio sources and the jets belong to the II type sources, as in the case of radio galaxy PKS1549-79 (Tadhunter et al. 2001).

The typical members of type 2 radio sources family are gigahertz-peaked spectrum sources (GPS), radio emission from GPS sources being produced by the expanding jets (Nagar et al. 2002).

Radio spectrum of galaxy M31 is described by the law (14) in the wavelength range of 20 cm to 170 cm (Sharov 1982).

Radio emission of diffuse nebulae is similar to that of planetary nebulae (Lang 1978)

## 8. CONCLUSIONS

Thus we conclude that the radio spectra of known types of extended radio sources can be described by the theory of incoherent induced radiation for gaseous disk, which was developed above.

The possibility to apply the model of gaseous disk to gaseous nebulae and active galactic nuclei is justified by the similarity of optical emission spectra. Besides, there is a non-uniform shift of the spectral lines in the spectra of quasars and radio galaxies, similar to those in the spectra of planetary nebulae.

The thermal origin of emission in the case of active galactic nuclei is supported by the recent observation of the [OIII] lines polarization in the spectrum of radio galaxy PKS1549-79 coincident with the polarization of optical continuum (Tadhunter et al. 2001).

The author is grateful to D.A.Kompaneets, Y.Y.Kovalev, V.A.Kostyuchenko, V.N.Lukash, V.V.Ovcharov, A.V.Postnikov, V.Semenov, and B.E.Stern for useful discussions.

-----------------------------------------------------------

Ulvestad, J.S. & Ho, L.C. 2001, ApJ, 562, L133

**Figure 1** Temperature as a function of height in Venus atmosphere . The data upon the pressure and the temperature in the lower layer of Venus atmosphere[10] have been used to start

up the transformation of the relation between the brightness temperature and wavelength into the temperature section across the atmosphere with the help of equations (1) and (5). Venus atmosphere is assumed to be consisting of carbon dioxide[10] .

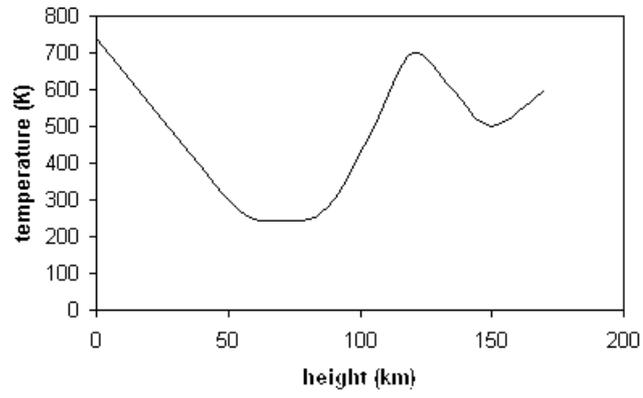

FIG.1. Temperature as a function of height in Venus atmosphere.